\documentclass[superscriptaddress,aps,preprintnumbers,prd,nofootinbib,preprint]{revtex4-1}

\usepackage{graphicx} 
\usepackage{amsmath,amssymb,bm}
\usepackage{xcolor}
\usepackage[colorlinks,allcolors=blue]{hyperref} 
\usepackage{mathtools}

\begin{document}
\preprint{TU-1301, RIKEN-iTHEMS-Report-26, OCU-PHYS 623, AP-GR 210}
\title{
Multi-field oscillons/I-balls in the
Friedberg-Lee-Sirlin model}

\author{Kai Murai}
\affiliation{RIKEN Center for Interdisciplinary Theoretical and Mathematical Sciences (iTHEMS), RIKEN, Wako 351-0198, Japan}
\affiliation{Department of Physics, Tohoku University, 
Sendai, Miyagi 980-8578, Japan} 
\author{Tatsuya Ogawa}
\affiliation{Department of Physics, Tohoku University, 
Sendai, Miyagi 980-8578, Japan} 
\affiliation{Tokyo Denki University, Tokyo 120-8551, Japan} 
\affiliation{Osaka Central Advanced Mathematical Institute, Osaka Metropolitan University,
Osaka 558-8585, Japan} 
\author{Fuminobu Takahashi}
\affiliation{Department of Physics, Tohoku University, 
Sendai, Miyagi 980-8578, Japan} 
\affiliation{Kavli IPMU (WPI), UTIAS, University of Tokyo, Kashiwa 277-8583, Japan}

\begin{abstract}
We study oscillon/I-ball solutions in a real scalar version of the Friedberg-Lee-Sirlin (FLS) model. Using the multiple-scale analysis, we derive the conditions for oscillon solutions and explore multi-field oscillon configurations. In these configurations, the two fields form co-located oscillons that oscillate with frequencies set by their respective masses. These multi-field oscillons can be viewed as a bound state of two oscillons due to attractive interactions between the fields.
We confirm these analytical predictions through numerical lattice calculations. This work extends the standard picture of single-field oscillons and may be relevant for cosmological scenarios involving multiple interacting real scalar fields.
\end{abstract}

\maketitle

\section{Introduction}
\label{sec: intro}
Oscillons/I-balls are non-topological quasi-solitons formed by real scalar fields~\cite{Bogolyubsky:1976yu,Gleiser:1993pt,Copeland:1995fq,Kasuya:2002zs}.%
\footnote{
    We use the terms oscillons and I-balls interchangeably. The term I-ball is used to emphasize that the stability arises from the adiabatic invariant~\cite{Kasuya:2002zs}.
}
Unlike Q-balls~\cite{Coleman:1985ki}, whose stability is guaranteed by the conservation of global U(1) charge, oscillons do not have an apparent symmetry in their Lagrangian. However, their dynamics exhibit an approximate conservation of an adiabatic invariant, namely the particle number~\cite{Kasuya:2002zs,Mukaida:2014oza,Kawasaki:2015vga}, and oscillons/I-balls represent field configurations that minimize energy under a given adiabatic invariant.
These structures are sometimes referred to as axitons, oscillatons, or breathers, depending on the context. In recent years, there has been extensive research on the generation and evolution of oscillons/I-balls in axion fields and inflaton fields~\cite{Amin:2011hj,Lozanov:2017hjm,Hiramatsu:2020obh,Kawasaki:2019czd,Olle:2019kbo,Vaquero:2018tib,Kawasaki:2020jnw,Kawasaki:2020tbo,Cyncynates:2021xzw,Kawasaki:2021poa,Cyncynates:2022wlq} as well as their decay processes~\cite{Gleiser:2008ty,Fodor:2008du,Hertzberg:2010yz,Salmi:2012ta,Mukaida:2016hwd,Ibe:2019vyo,Ibe:2019lzv,Zhang:2020bec,Cyncynates:2021rtf,Imagawa:2021sxt,Nakayama:2023jhg}. 

Similar to oscillons/I-balls, there exists a non-topological soliton called a Q-ball~\cite{Coleman:1985ki}, whose stability is ensured by the conservation of a global U(1) charge. In fact, before Coleman proposed Q-balls~\cite{Coleman:1985ki}, Friedberg, Lee, and Sirlin introduced a model (the FLS model) involving one complex scalar field and one real scalar field~\cite{Friedberg:1976me}, which admits Q-ball solutions and can be considered as a type of UV completion for a class of models realizing Q-balls. The FLS model has recently attracted renewed interest and has been revisited~\cite{Heeck:2023idx,Kim:2023zvf,Kim:2024vam,Hamada:2024pbs,Jaramillo:2024cus}.

Inspired by the FLS model, in this paper, we investigate whether oscillon/I-ball solutions can exist in a real-scalar version of the FLS model.
Oscillons have typically been studied in the context of a single real scalar field. 
In the real FLS model, the system can be reduced to an effective single-field theory in the limits where one scalar $\varphi$ is much heavier or much lighter than the other scalar $\psi$.
In both limits, we find oscillon solutions composed of the lighter scalar field.
This observation motivates us to study the full two-field system and examine whether oscillon solutions involving both scalar fields exist away from these single-field limits.

However, the existence and properties of oscillons in systems with two or more interacting real scalar fields remain an interesting open problem. 
Using a small-amplitude expansion based on multiple-scale analysis, we analytically derive the conditions under which oscillon solutions exist in the real FLS model. We further show the existence of true multi-field oscillon solutions, in which two fields simultaneously form an oscillon at the same location. These two-field oscillons are distinct from Q-balls formed by complex scalar fields in that the two scalar fields oscillate with different periods. In particular, the oscillation frequency of each component is approximately set by the mass of the corresponding scalar field.
We then demonstrate with lattice simulations that multi-field oscillons are produced dynamically, and that their properties are consistent with our analytic predictions. These multi-field oscillons can be interpreted as bound states of two oscillons held together by an attractive interaction.

We briefly comment on related work in the literature. Ref.~\cite{Gleiser:2011xj} found two-field oscillons in a closely related two-scalar model and identified ground-state and excited-state configurations. Their explicit numerical solutions were presented for a benchmark in which the small-amplitude oscillation frequencies of the two fields around the vacuum are nearly degenerate. We will see that their excited-state solution can be naturally interpreted as a driven configuration in which one field oscillates at approximately twice the frequency of the other.
In our multiple-scale analysis for two fields, it corresponds to an oscillon solution in which one field vanishes at the leading order.
Thus, in some sense, their excited solution is effectively a single-field oscillon.
Oscillons in two-field systems have also been studied in various contexts, such as models symmetric under the exchange of the two fields~\cite{Hawley:2002zn,Correa:2019jaa,VanDissel:2020umg}, the Abelian-Higgs model~\cite{Achilleos:2013zpa,Diakonos:2014hra}, and models in which an additional field affects oscillons of one field~\cite{Kawasaki:2013awa,Antusch:2015ziz,Shafi:2024jig,Li:2025ioq}.

In contrast, we study two-field oscillons in the real FLS model for a broad range of mass ratios. Using a multiple-scale analysis, we analytically derive their profiles and show that they are in good agreement with our lattice simulation results. Rather than oscillating with a single common frequency, the two fields oscillate predominantly with frequencies set by their own masses, while their respective adiabatic invariants are well conserved, suggesting stabilization as a multi-field I-ball. This in turn supports an interpretation of the configuration as a bound state of two oscillons held together by their mutual attraction. We also find solutions with markedly non-Gaussian profiles, including configurations with extended tails and configurations in which the two constituent oscillons have substantially different radii. In such cases, the commonly assumed picture of a single coherent phase across the whole oscillon does not hold. Nevertheless, the object remains localized and long-lived, and its energy profile is still reproduced well by the multiple-scale analysis.

The remainder of this paper is organized as follows. In Sec.~\ref{sec:model}, we introduce the original FLS model and its real-scalar version. In Sec.~\ref{sec:oscillon}, we analytically study single-field oscillons in the real FLS model using a multiple-scale analysis. In Sec.~\ref{sec: two timing analysis}, we extend the multiple-scale analysis to multi-field oscillons and derive their properties and existence conditions. In Sec.~\ref{sec:simulation}, we present results from lattice simulations and compare them with the analytic predictions. Finally, Sec.~\ref{sec: summary} is devoted to conclusions and discussion.

\section{Friedberg-Lee-Sirlin model}
\label{sec:model}

In this section, we introduce the model studied in this paper.
We first review the original Friedberg-Lee-Sirlin (FLS) model, which contains two scalar fields: one real and one complex.
We then define the real-scalar version that will be the main focus of our analysis.

\subsection{Original FLS model}
\label{subsec:original_model}

As an early study of spherically symmetric non-topological solitons, Friedberg, Lee, and Sirlin proposed the model~\cite{Friedberg:1976me}
\begin{align}
    \mathcal{L}
    =
    -\frac{1}{2} \partial_\mu \phi \partial^\mu \phi
    - \partial_{\mu} \psi^{\ast}\partial^{\mu}\psi
    - \frac{\lambda}{4}(\phi^2- \eta^2)^2
    - \kappa |\psi|^2 \phi^2
    \ ,
    \label{eq:original_FLS_Lagrangian}
\end{align}
where $\phi$ and $\psi$ are a real and a complex scalar field, respectively, and $\lambda$ and $\kappa$ are coupling constants.
The Lagrangian is invariant under the $Z_2$ symmetry of $\phi$, $\phi \to -\phi$, and the global U(1) symmetry of $\psi$, $\psi \to e^{i\theta}\psi$.
In the vacuum, $\phi$ acquires a nonzero vacuum expectation value (VEV), spontaneously breaking the $Z_2$ symmetry, while the U(1) symmetry remains unbroken.
The VEVs are
\begin{align}
    \langle \phi \rangle = \pm \eta
    \ , \quad 
    \langle \psi \rangle = 0
    \ .
    \label{eq:VEVs_phi_psi}
\end{align}
Since we are not concerned with topological defects in this paper, we choose the vacuum $\langle \phi \rangle = \eta > 0$ without loss of generality, and study the dynamics around it.
Expanding around the VEVs, the masses of $\phi$ and $\psi$ are
$m_\phi = \sqrt{2\lambda}\,\eta$ and $m_\psi = \sqrt{\kappa}\,\eta$, respectively.

When $m_\phi \gg m_\psi$, the field $\phi$ is heavy and can be eliminated in an effective description at energies well below $m_\phi$.
In this regime, the resulting effective potential for $\psi$ can become shallower than quadratic over a relevant field range, which is the basic ingredient for the existence of Q-ball solutions~\cite{Coleman:1985ki}.
If one replaces $\psi$ by a real scalar field, the same mechanism suggests the possibility of oscillon/I-ball solutions built from that real field.

In contrast, when $m_\phi \ll m_\psi$, the structure of the potential is similar to that in hybrid inflation models.
Near the minima of the double-well potential, the potential for $\phi$ is shallower than quadratic.
Numerical simulations have shown that oscillons of $\phi$ can form in hybrid inflation~\cite{McDonald:2001iv,Copeland:2002ku,Broadhead:2005hn,Gleiser:2011xj}.

\subsection{Real scalar FLS model}
\label{subsec:real_model}

The original FLS model in Eq.~\eqref{eq:original_FLS_Lagrangian} possesses a global U(1) symmetry acting on the complex field $\psi$.
Here, we instead take $\psi$ to be a real scalar field.
The Lagrangian is then
\begin{align}
    \mathcal{L}
    =
    -\frac{1}{2} \partial_{\mu}\phi \partial^{\mu}\phi
    -\frac{1}{2} \partial_{\mu}\psi \partial^{\mu}\psi
    - V(\phi,\psi)
    \ ,
    \label{eq:real_FLS_Lagrangian}
\end{align}
with the scalar potential
\begin{align}
    V(\phi,\psi)
    = \frac{\lambda}{4}(\phi^2-\eta^2)^2
    + \kappa \psi^2 \phi^2
    \ .
    \label{eq:scalar_potential_phi_psi}
\end{align}
The vacuum structure is the same as in Eq.~\eqref{eq:VEVs_phi_psi}, namely $\langle \phi\rangle = \pm \eta$ and $\langle \psi\rangle = 0$.
Expanding around $\langle \phi\rangle=\eta$, the masses are
\begin{align}
    m_\phi = \sqrt{2\lambda}\,\eta
    \ , \qquad
    m_\psi = \sqrt{2\kappa}\,\eta
    \ .
    \label{eq:masses_real_FLS}
\end{align}

To study the dynamics around the vacuum, we write $\phi=\eta+\varphi$.
The potential becomes
\begin{align}
    V(\varphi,\psi)
    &=
    \frac{m_\phi^2}{2}\,\varphi^2
    + \sqrt{\frac{\lambda}{2}}\, m_\phi\,\varphi^3
    + \frac{\lambda}{4}\,\varphi^4
    + \frac{m_\psi^2}{2}\,\psi^2
    + \sqrt{2\kappa}\, m_\psi\, \varphi\,\psi^2
    + \kappa\,\varphi^2\psi^2
    \ .
    \label{eq:expanded_potential_real_FLS}
\end{align}
As we will show in the next section, depending on the hierarchy between $m_\phi$ and $m_\psi$, this model admits oscillon solutions predominantly composed of either $\varphi$ or $\psi$.
This is closely related to the fact that, in the original FLS model, the complex field $\psi$ supports Q-ball solutions, while the real field $\phi$ can form oscillons.

\section{Oscillon/I-ball solutions for a single field}
\label{sec:oscillon}

In this section, we analyze oscillon/I-ball solutions in the real-scalar FLS model.
We focus on a regime with a mass hierarchy between $\varphi$ and $\psi$, in which the dynamics effectively reduces to a single-field system and oscillons are predominantly formed by one field.

\subsection{Single-field oscillon/I-ball solution}
\label{subsec: oscillon}

Before investigating the specific setup, we briefly review the oscillon/I-ball solutions in a single real field theory.
Here, we consider the Lagrangian of a real scalar field $\chi$ given by
\begin{align}
    \mathcal{L} 
    =
    -\frac{1}{2} (\partial_{\mu} \chi)(\partial^{\mu} \chi)-V(\chi)
    \ .
\end{align}
Then, $\chi$ follows the equation of motion (EOM),
\begin{align}
    \frac{\partial^2 \chi}{\partial t^2}  - \nabla^2 \chi + \frac{\partial V(\chi)}{\partial \chi}
    =
    0
    \ .
\end{align}
Since the oscillons are known to have spherically symmetric configurations, we assume $\chi(\bm{x}) = \chi(r)$ with $r \equiv |\bm{x}|$ in the following.
In addition, we expand $V(\chi)$ as
\begin{align}
    V(\chi) 
    =
    \frac{1}{2} m^2 \chi^2 + \frac{\lambda_3}{3!} \chi^3 + \frac{\lambda_4}{4!} \chi^4 
    + \mathcal{O}(\chi^5)
    \ .
\end{align}
Then, the EOM becomes 
\begin{align}
    \frac{\partial^2 \chi}{\partial t^2}
    - \frac{\partial^2 \chi}{\partial r^2} 
    - \frac{d-1}{r}\frac{\partial \chi}{\partial r} 
    + m^2 \chi + \frac{\lambda_3}{2} \chi^2 + \frac{\lambda_4}{6} \chi^3
    + \mathcal{O}(\chi^4)
    =
    0
    \ ,
\end{align}
where $d$ is the spatial dimension.

Here, we derive the spatial configurations of oscillons and discuss the conditions for the existence of oscillon solutions using the small-amplitude expansion based on multiple-scale analysis.
Since oscillons have angular frequencies slightly smaller than the mass, we separate the time dependence of oscillon solutions into two timescales: one is set by the inverse mass, and the other by the small difference between the oscillation frequency and the mass.
While the former is described by the cosmic time $t$, the latter can be described by a rescaled time coordinate $\tau$,
\begin{align}
    \tau \equiv \epsilon^2 t
    \ ,
\end{align}
where $\epsilon \ll 1$. 
A function of $\tau$ therefore evolves slowly with respect to $t$. For example, $\cos(\tau) = \cos(\epsilon^2 t)$ varies on the slow timescale.
In addition, for the adiabatic charge to be conserved~\cite{Kasuya:2002zs, Kawasaki:2015vga}, the spatial scale of oscillon/I-ball solutions must be much larger than the inverse mass scale.
Thus, we also introduce a rescaled spatial variable
\begin{align}
    \rho \equiv  \epsilon r
    \ ,
\end{align}
to express the gradual spatial dependence of oscillon configurations.
Here, we fixed the dependences of $\tau \propto \epsilon^2$ and $\rho \propto \epsilon$ so that the lowest order of $\epsilon$ from the time derivative and spatial derivatives matches in the EOMs, as we will see below.

Here, we comment on different formulations of the multiple-scale analysis. 
In the oscillon literature, a rescaled time variable, $\sqrt{1-\epsilon^2}t$, is often used instead of introducing two time variables, $t$ and $\tau$ (see, e.g., Refs.~\cite{Kichenassamy:1991vlk,Fodor:2008es,Hertzberg:2010yz}).
In the single-field case, these two formulations lead to the same equation for the oscillon profile.
However, introducing the slow time variable $\tau$ is useful in multi-field systems, where different fields can acquire different corrections to their oscillation frequencies.
We therefore introduce $\tau$ as in Ref.~\cite{VanDissel:2020umg} and extend this formulation to the two-field system in the following section.%
\footnote{%
    While the use of $\tau$ is analogous to that in Ref.~\cite{VanDissel:2020umg}, their model is symmetric under the exchange of the two scalar fields, and the introduction of $\tau$ is not essential there.
}

Using these new variables, we represent the solution as functions of $(t, \tau, \rho)$, although $t$ and $\tau$ are not originally independent.
Then, the time derivative is replaced as $\partial_t \to \partial_t + \epsilon^2 \partial_\tau$.
As a result, we rewrite the EOMs as
\begin{align}
    \partial_t^2 \chi 
    + 2 \epsilon^2 \partial_t \partial_{\tau} \chi
    - \epsilon^2 \left( \partial_{\rho}^2 + \frac{d-1}{\rho}\partial_{\rho} \right) \chi
    + m^2 \chi 
    + \frac{\lambda_3}{2} \chi^2 
    + \frac{\lambda_4}{6} \chi^3
    + \mathcal{O}( \epsilon^4 )
    &=
    0
    \ .
    \label{eq:EOM_chi_twotiming}
\end{align}

In addition, we also assume that $\chi$ has a small amplitude so that the potential is dominated by the quadratic term~\cite{Kasuya:2002zs} and the effects of nonlinear interaction terms can be incorporated perturbatively.
Thus, we expand the solutions in the form of 
\begin{align}
    \chi(t, \tau, \rho)
    =
    \sum_{n=1} \epsilon^n \chi_n(t, \tau, \rho)
    \ ,
    \label{eq:expand_chi}
\end{align}
Substituting Eq.~\eqref{eq:expand_chi} into Eq.~\eqref{eq:EOM_chi_twotiming}, we obtain
\begin{align}
    &\sum_{n=1} \left[ 
        \epsilon^n \partial_t^2 \chi_n
        + 2 \epsilon^{2+n} \partial_t \partial_{\tau} \chi_n
        - \epsilon^{2+n} \left( 
            \partial_{\rho}^2
            + \frac{d-1}{\rho} \partial_{\rho}
        \right) \chi_n
    \right]
    \nonumber \\
    & + m^2 \sum_{n=1} \epsilon^{n} \chi_n
    + \frac{\lambda_3}{2} \left( \sum_{n=1} \epsilon^{n} \chi_n \right)^2
    + \frac{\lambda_4}{6} \left( \sum_{n=1} \epsilon^{n} \chi_n \right)^3
    =
    0
    \ .
\end{align}
In the following, we solve these EOMs order by order.

\subsubsection{The lowest order \texorpdfstring{$\mathcal{O}(\epsilon)$}{}}

The EOMs for the lowest order $\mathcal{O(\epsilon)}$ are given by
\begin{align}
    \partial_t^2 \chi_1 + m^2 \chi_1
    &=
    0
    \ .
    \label{eq:EOM_chi_order_1}
\end{align}
The solution is 
\begin{align}
    \chi_1(t, \tau(t), \rho)
    &=
    \mathrm{Re}[C(\tau,\rho)e^{-imt}]
    =
    \frac{1}{2}\left[ 
        Ce^{-imt} + C^{\ast}e^{imt} 
    \right]
    \ ,
    \label{eq:solution_chi_order_1}
\end{align}
where $C(\tau,\rho)$ is a complex function representing the oscillation amplitudes for $\chi$.

\subsubsection{The second order \texorpdfstring{$\mathcal{O}(\epsilon^2)$}{}}

For the next order $\mathcal{O}(\epsilon^2)$, the EOMs are given by
\begin{align}
    \partial_t^2 \chi_2 
    + m^2 \chi_2 
    + \frac{\lambda_3}{2} \chi_1^2
    &=
    0
    \ , 
    \label{eq:EOM_chi_order_2}
\end{align}
The EOM shows a new oscillatory term arising from the nonlinear interactions, dependent on $\chi_1$.
By substituting them in Eq.~\eqref{eq:solution_chi_order_1}, we can rewrite the EOM as
\begin{align}
    \partial_t^2{\chi}_2 + m^2 \chi_2
    &=
    -\frac{\lambda_3}{8}
    \left\{
        C^2 e^{-2imt} + 2|C|^2 
        + C^{\ast 2}e^{2imt}
    \right\}
    \ .
\end{align}
We can solve these equations to obtain
\begin{align}
    \chi_2
    &=
    \frac{\lambda_3}{24 m^2}
    \left(
        C^2 e^{-2imt} 
        - 6|C|^2 
        + C^{\ast 2}e^{2imt}
    \right)
        + c_1 \cos(mt)
        + c_2 \sin(mt) 
    \ ,
\end{align}
where $c_1$ and $c_2$ are integration constants with respect to the fast time $t$ and may in general depend on $\tau$ and $\rho$.
Since $\chi_2$ contributes to $\chi$ at $\mathcal{O}(\epsilon^2)$ and is subleading compared to the $\chi_1$ term for $\epsilon \ll 1$, we do not explicitly derive the equations that $c_1$ and $c_2$ follow.
They can be obtained from the EOMs for the fourth order $\mathcal{O}(\epsilon^4)$, analogously to the derivation of the equation for $C$ below.

\subsubsection{The third order \texorpdfstring{$\mathcal{O}(\epsilon^3)$}{}}

The EOMs for the third order $\mathcal{O}(\epsilon^3)$ are given by
\begin{align}
    \partial_t^2 \chi_3 + m^2 \chi_3
    &=
    \left( \partial_{\rho}^2+\frac{d-1}{\rho}\partial_{\rho} \right)\chi_1
    - 2 \partial_t \partial_{\tau} \chi_1
    - \lambda_3 \chi_1 \chi_2
    - \frac{\lambda_4}{6} \chi_1^3
    \ .
\end{align}
Substituting $\chi_1$ and $\chi_2$, we find that the right-hand side (RHS) includes the terms depending on $t$ as $e^{j \times imt}$ with $j = -3,-2,\ldots, 3$.
Among them, the terms proportional to $e^{-i m t}$ and $e^{i m t}$ are secular terms.
If secular terms exist, $\chi_3$ would grow over time $t$.
To ensure that $\chi_3$ remains bounded and physically consistent, these secular terms must be eliminated from the RHS.

Removing the secular terms results in the following equations governing the time evolution of the amplitude function $C(\tau,\rho)$:
\begin{align}
    &\left( \partial_{\rho}^2+\frac{d-1}{\rho}\partial_{\rho} \right) C
    +2im \partial_{\tau} C
    + 
    \left(\frac{5\lambda_3^2}{24m^2} - \frac{\lambda_4}{8} \right)C |C|^2  
    =
    0
    \ .
    \label{eq:EOM_C}
\end{align}
The derived equation determines the amplitude function, $C$, which describes the spatial profiles and the slow time evolution of the oscillons, capturing the influence of nonlinear interactions.

Since oscillons are localized structures that oscillate in time, we assume solutions of the separable form
\begin{align}
    C(\tau,\rho)
    =
    c(\rho)e^{i\omega \tau}
    \ ,
    \label{eq: C and c}
\end{align}
where $\omega$ is a constant.
By substituting this ansatz into Eq.~\eqref{eq:EOM_C}, we obtain the differential equation for the radial profile $c(\rho)$ as
\begin{align}
    & \frac{\mathrm{d}^2 c}{\mathrm{d}\rho^2}
    + \frac{d-1}{\rho} \frac{\mathrm{d}c}{\mathrm{d}\rho}
    - 2 \omega m c
    + \left(\frac{5\lambda_3^2}{24m^2} - \frac{\lambda_4}{8} \right) c |c|^2
    =
    0
    \ .
    \label{eq:EOM_c}
\end{align}
To ensure a regular solution with finite energy, we impose the boundary conditions, 
\begin{equation}
\begin{gathered}
    \frac{dc}{d\rho}(\rho = 0) 
    =
    0
    \ ,
    \\
	c(\rho \to \infty)
    =
    0
    \ .
    \label{eq: c boundary}
\end{gathered}
\end{equation}
While $c$ is complex in general, its complex phase becomes independent of $\rho$ under these boundary conditions.
Thus, we set $c$ to be real by shifting the origin of $\tau$ in Eq.~\eqref{eq: C and c}.
Then, Eq.~\eqref{eq:EOM_c} can be rewritten using an effective potential $V_\mathrm{eff}$ as 
\begin{align}
    & \frac{\mathrm{d}^2c}{\mathrm{d}\rho^2}
    + \frac{d-1}{\rho} \frac{\mathrm{d}c}{\mathrm{d}\rho}
    + \frac{\partial V_\mathrm{eff}}{\partial c}
    =
    0
    \ ,
\end{align}
with 
\begin{align}
    V_\mathrm{eff}(c)
    \equiv &
    - \omega m c^2
    +
    \frac{1}{32}
    \left(\frac{5\lambda_3^2}{3m^2} - \lambda_4 \right) c^4
    \ .
    \label{eq: V_eff}
\end{align}
Here, the presence of the cubic coupling $\lambda_3$ effectively induces a negative quartic coupling in the effective potential.
This implies that the exchange of a scalar particle induces an attractive force.

Equation~\eqref{eq:EOM_c} can be interpreted analogously to the time evolution of a homogeneous scalar field in an expanding universe with an effective potential of $V_\mathrm{eff}(c)$.
Therefore, $V_\mathrm{eff}$ must have a positive quartic term for solutions satisfying the boundary conditions~\eqref{eq: c boundary} to exist.
In other words, the condition is
\begin{align}
    \lambda_4 
    <
    \frac{5 \lambda_3^2}{3 m^2}
    \ .
\end{align}

By solving Eq.~\eqref{eq:EOM_c} under these boundary conditions, we derive an oscillon configuration in the form of
\begin{align}
    \chi(t,r)
    &=
    \epsilon c(\epsilon r) \cos \left[ 
        \left( m - \epsilon^2\omega \right) t
    \right]
    + \mathcal{O}(\epsilon^2)
    \ .
\end{align}
This solution represents a spatially localized oscillon with corrections up to order $\mathcal{O}(\epsilon^2)$, where $\chi$ oscillates with frequencies shifted by small corrections.

Here, we comment on the parameter of physically distinct oscillon solutions.
While $m$, $\lambda_3$, and $\lambda_4$ are model parameters, $\omega$ is introduced in deriving the differential equation for $c$, i.e., Eq.~\eqref{eq:EOM_c}.
Thus, $\omega$ works as a parameter characterizing the profile function $c$.
However, in specifying the full oscillon configuration, $\chi(t,r)$, there is another parameter $\epsilon$, and actually $\epsilon$ and $\omega$ are degenerate as a parameter characterizing the full oscillon configuration $\chi(t,r)$.
To see this explicitly, let us denote a solution of $c$ for a given value of $\omega$ by $c(\rho; \omega)$.
From the structure of Eq.~\eqref{eq:EOM_c}, we find 
\begin{align}
    c(\rho; \alpha^2\omega) = \alpha c(\alpha \rho; \omega)
    \ ,
\end{align}
where $\alpha > 0$ is a rescaling parameter.
Using this relation, we see that $\chi(t,r)$ obtained with the parameters $(\alpha^2 \omega, \epsilon)$ is identical to that obtained with $(\omega, \alpha \epsilon)$.
In other words, the difference in $\omega$ can be compensated by a rescaling of $\epsilon$, and hence the oscillon configuration is characterized only by the corresponding combined scale.
Therefore, it is sufficient to solve Eq.~\eqref{eq:EOM_c} for a fixed value of $\omega$, for example $\omega = m_\phi$, and vary $\epsilon$ to obtain the full family of physically distinct oscillon configurations.

\subsection{Effective single-field limit in the real FLS model}

\subsubsection{\texorpdfstring{$\phi$}{}-oscillon for \texorpdfstring{$m_\psi \gg m_\phi$}{}}
\label{subsec: heavy psi}

If the masses of $\phi$ and $\psi$ are hierarchical, the system can be reduced to an effective single-field theory for the lighter field.
Here we consider the limit  $m_\psi \gg m_\phi$ and apply the analysis above to the resulting effective theory.

When $\psi$ is much heavier than $\phi$, $\psi$ stays at its  minimum, $\psi = 0$.
We then obtain the effective potential for $\phi$,
\begin{align}
    V_{\psi\text{-int}}(\phi)
    =
    \frac{m_\phi^2}{2} \varphi^2
    + \sqrt{\frac{\lambda}{2}} m_\phi \varphi^3
    + \frac{\lambda}{4} \varphi^4    
    \ .
\end{align}
Using this potential, for the oscillon ansatz
\begin{align}
    \phi(t,r)
    =
    \epsilon a(\epsilon r) 
    \cos \left[ \left( m_\phi - \epsilon^2 \omega_\varphi \right) t \right]
    +
    \mathcal{O}(\epsilon^2)
    \ ,
\end{align}
we obtain the effective potential for $a$
\begin{align}
    V_\mathrm{eff}(a)
    =
    - \omega_\varphi m_\phi a^2
  + \frac{3\lambda}{4} a^4    
    \ .
\end{align}
Thus,  oscillon solutions exist for $\lambda > 0$.
The equation for the spatial configuration is given by 
\begin{align}
    \frac{\mathrm{d}^2 a}{\mathrm{d}\rho^2} 
    + \frac{d-1}{\rho} \frac{\mathrm{d} a}{\mathrm{d}\rho} 
    - 2 \omega_\varphi m_{\phi} a
    + 3\lambda a^3 
    = 0 
    \ .
    \label{eq: a EoM integrate psi}
\end{align}

\subsubsection{\texorpdfstring{$\psi$}{}-oscillon for \texorpdfstring{$m_\phi \gg m_\psi$}{}}
\label{subsec: heavy phi}

When $\phi$ is much heavier than $\psi$, we can integrate out $\phi$ by assuming that it adiabatically follows the potential minimum,
\begin{align}
    \phi^2 = \eta^2 - \frac{2\kappa}{\lambda} \psi^2
    \ ,
\end{align}
which is obtained from $\partial V(\phi,\psi)/\partial \phi = 0$.
Substituting this relation into Eq.~\eqref{eq:scalar_potential_phi_psi}, we obtain the effective potential for $\psi$,
\begin{align}
    V_{\phi\text{-int}}(\psi)
    =
    \frac{m_\psi^2}{2} \psi^2
    - \frac{\kappa^2}{\lambda} \psi^4    
    \ .
\end{align}
For the oscillon ansatz,
\begin{align}
    \psi(t,r)
    =
    \epsilon b(\epsilon r) 
    \cos \left[ \left( m_\psi - \epsilon^2 \omega_\psi \right) t \right]
    +
    \mathcal{O}(\epsilon^2)
    \ ,
\end{align}
we find
the effective potential,
\begin{align}
    V_\mathrm{eff}(b)
    =
    - \omega_\psi m_\psi b^2
    + \frac{3\kappa^2}{4\lambda} b^4    
    \ .
\end{align}
Therefore, oscillon solutions exist for $\lambda > 0$ and $\kappa \neq 0$.
The equation for the spatial configuration is 
\begin{align}
    \frac{\mathrm{d}^2 b}{\mathrm{d}\rho^2} 
    + \frac{d-1}{\rho} \frac{\mathrm{d}b}{\mathrm{d}\rho} 
    - 2 \omega_\psi m_{\psi} b
    + \frac{3\kappa^2}{\lambda} b^3 
    = 0 
    \ .
    \label{eq: b EoM integrate phi}
\end{align}
Notice that, if we set $\kappa = 0$, the effective potential for $\psi$ reduces to a quadratic form, which does not support oscillon solutions.
In this sense, the coupling between $\phi$ and $\psi$ is essential for the existence of $\psi$-oscillon solutions, even in the heavy $\phi$ limit.

\begin{figure}
    \centering
    \includegraphics[width=7.5cm]{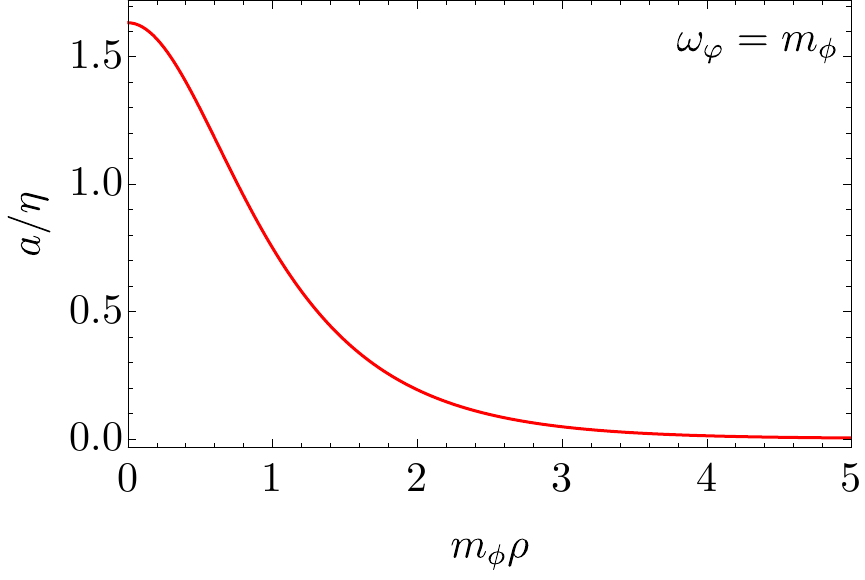}
    \hspace{5mm}
    \includegraphics[width=7.5cm]{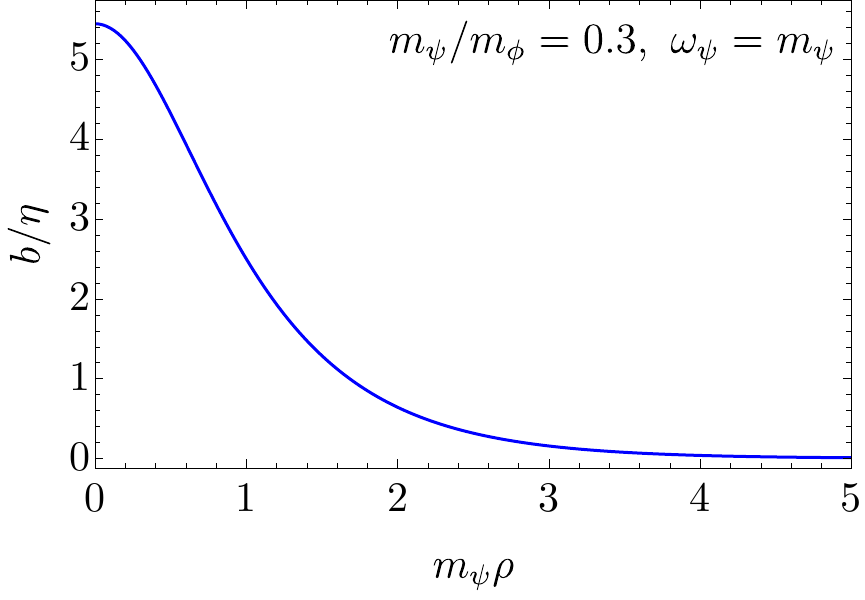}
    \caption{%
    Spatial profiles of single-field oscillons for $\phi$ (left) and $\psi$ (right).
    }
    \label{fig: single-field oscillon}
\end{figure}

We can obtain the spatial profile functions, $a$ and $b$ by solving the corresponding differential equations~\eqref{eq: a EoM integrate psi} and \eqref{eq: b EoM integrate phi}.
We show the numerical spatial profiles for oscillons of $\phi$ and $\psi$ in Fig.~\ref{fig: single-field oscillon}.
For both panels, we take $d = 1$. In the left panel, corresponding to the $\phi$-oscillon, we set $\omega_\varphi = m_\phi$. In the right panel, corresponding to the $\psi$-oscillon, we set $\omega_\psi = m_\psi$, and $m_\psi/m_\phi = 0.3$.
The real scalar fields $\phi$ and $\psi$ are localized within a finite spatial region and oscillate with constant angular frequencies.

\section{Multi-field oscillons in multiple-scale analysis}
\label{sec: two timing analysis}

In the previous section, we showed that in the effective single-field limits, oscillon solutions exist for either $\phi$ or $\psi$, depending on the mass hierarchy.
In this section, we turn to the case where both fields participate and form a localized, long-lived configuration, which we call a multi-field oscillon.
To this end, we extend the multiple-scale analysis to the two-field system and derive the equations governing the spatial profiles of multi-field oscillons.

From here on, we normalize dimensionful variables by the symmetry-breaking scale $\eta$ and the mass of $\phi$:
\begin{equation}
\begin{gathered}
    t\to \tilde{t} \equiv \sqrt{2\lambda} \eta t = m_\phi t
    \ , \quad 
    r\to \tilde{r} \equiv \sqrt{2\lambda} \eta r = m_\phi r
    \ , 
    \\
    \phi\to \tilde{\phi} \equiv \phi/\eta
    \ , \quad 
    \psi\to \tilde{\psi} \equiv \psi/\eta
    \ .
    \label{eq: dimless quantities}
\end{gathered}
\end{equation}
In addition, we define the parameter
\begin{align}
    \mu \equiv \sqrt{\frac{\kappa}{\lambda}} 
    = \frac{m_\psi}{m_\phi}
    \ .
\end{align}
For simplicity, we drop tildes in the following.
Then, imposing spherical symmetry, the EOMs for $\varphi$ and $\psi$ are given by
\begin{align}
    - \partial_t^2 \varphi 
    + \left( \partial_r^2 + \frac{d-1}{r} \partial_r \right) \varphi
    - \frac{1}{2} \varphi ( \varphi + 2 ) ( \varphi+1 )
    - \mu^2 ( \varphi + 1 ) \psi^2
    &=
    0
    \ ,
    \\
    - \partial_t^2 \psi
    + \left( \partial_r^2 + \frac{d-1}{r} \partial_r \right) \psi
    - \mu^2 ( \varphi + 1 )^2 \psi
    &=
    0
    \ .
\end{align}

\subsection{Multiple-scale analysis}
\label{subsec: two timing analysis}

As in the single-field case, we introduce
\begin{align}
    \tau \equiv \epsilon^2 t
    \ , \quad 
    \rho \equiv  \epsilon r
    \ ,
\end{align}
with $\epsilon \ll 1$. 
We then rewrite the EOMs as
\begin{align}
    - \partial_t^2 \varphi 
    - 2 \epsilon^2 \partial_t \partial_{\tau} \varphi
    + \epsilon^2 \left( \partial_{\rho}^2 + \frac{d-1}{\rho}\partial_{\rho} \right) \varphi
    - \frac{1}{2} \varphi( \varphi + 2 )( \varphi + 1 )
    - \mu^2 ( \varphi + 1 ) \psi^2
    + \mathcal{O}( \epsilon^4 )
    &=
    0
    \ ,
    \label{eq:EOM_varphi}
    \\
    - \partial_t^2 \psi
    - 2 \epsilon^2 \partial_t \partial_{\tau} \psi
    + \epsilon^2 \left( \partial_{\rho}^2 + \frac{d-1}{\rho} \partial_{\rho} \right) \psi
    - \mu^2 ( \varphi + 1 )^2 \psi
    + \mathcal{O}( \epsilon^4 )
    &=
    0
    \ .
    \label{eq:EOM_psi}
\end{align}
By expanding the solutions as
\begin{align}
    \varphi(t, \tau, \rho)
    =
    \sum_{n=1} \epsilon^n \varphi_n(t, \tau, \rho)
    \ , \quad 
    \psi(t, \tau, \rho)
    =
    \sum_{n=1} \epsilon^n \psi_n(t, \tau, \rho)
    \ ,
    \label{eq:expand_phi_psi}
\end{align}
we obtain
\begin{align}
    &\sum_{n=1} \left[ 
        - \epsilon^n \partial_t^2 \varphi_n
        - 2 \epsilon^{2+n} \partial_t \partial_{\tau} \varphi_n
        + \epsilon^{2+n} \left( 
            \partial_{\rho}^2
            + \frac{d-1}{\rho} \partial_{\rho}
        \right) \varphi_n
    \right]
    \nonumber \\
    & - \frac{1}{2} \sum_{l=1} \epsilon^l \varphi_l
    \left( \sum_{m=1} \epsilon^m \varphi_m + 2 \right)
    \left( \sum_{n=1} \epsilon^n \varphi_n + 1 \right)
    - \mu^2 \left( \sum_{m=1} \epsilon^m \varphi_m + 1 \right)
    \left( \sum_{n=1} \epsilon^{n} \psi_n \right)^2
    =
    0
    \ ,
    \\
    &\sum_{n=1} \left[
        -\epsilon^n \partial_t^2 \psi_n
        - 2 \epsilon^{2+n} \partial_t \partial_{\tau} \psi_n
        + \epsilon^{2+n} \left(
            \partial_{\rho}^2
            + \frac{d-1}{\rho} \partial_{\rho}
        \right) \psi_n
    \right]
    - \mu^2 
    \left( \sum_{m=1} \epsilon^m \varphi_m + 1 \right)^2
    \sum_{n=1} \epsilon^{n} \psi_n 
    =
    0
    \ .
\end{align}
In the following, we solve these EOMs order by order in $\epsilon$.

\subsubsection{The lowest order \texorpdfstring{$\mathcal{O}(\epsilon)$}{}}

The EOMs at the lowest order $\mathcal{O(\epsilon)}$ are given by
\begin{align}
    - \partial_t^2 \varphi_1 - \varphi_1
    &=
    0
    \ , 
    \label{eq:EOM_varphi_order_1}
    \\
    - \partial_t^2 \psi_1 - \mu^2 \psi_1
    &=
    0
    \ .
    \label{eq:EOM_psi_order_1}
\end{align}
These EOMs are those of the harmonic oscillators and describe the leading oscillatory behavior of the multi-field oscillon.
The solutions to Eqs.~\eqref{eq:EOM_varphi_order_1} and \eqref{eq:EOM_psi_order_1} are
\begin{align}
    \varphi_1(t, \tau(t), \rho)
    &=
    \mathrm{Re}[A(\tau,\rho)e^{-i t}]
    =
    \frac{1}{2}\left[
        Ae^{-i t} + A^{\ast}e^{i t} 
    \right]
    \ ,
    \label{eq:solution_varphi_order_1}
    \\
    \psi_1(t, \tau(t), \rho)
    &=
    \mathrm{Re}[B(\tau,\rho)e^{-i\mu t}]
    =
    \frac{1}{2}\left[
        Be^{-i\mu t} + B^{\ast}e^{i\mu t} 
    \right]
    \ ,
    \label{eq:solution_psi_order_1}
\end{align}
where $A(\tau,\rho)$ and $B(\tau,\rho)$ are complex functions of the slow time variable $\tau$ and the spatial variable $\rho$. 
They represent the oscillation amplitudes for the fields $\varphi$ and $\psi$, respectively, and vary on timescales much longer than the rapid oscillations in $t$.

\subsubsection{The second order \texorpdfstring{$\mathcal{O}(\epsilon^2)$}{}}

At the next order, $\mathcal{O}(\epsilon^2)$, the EOMs are
\begin{align}
    - \partial_t^2 \varphi_2 - \varphi_2 
    - \frac{3}{2} \varphi_1^2 - \mu^2 \psi_1^2
    &=
    0
    \ , 
    \label{eq:EOM_varphi_order_2}
    \\
    - \partial_t^2 \psi_2 - \mu^2 \psi_2 - 2 \mu^2 \varphi_1\psi_1
    &=
    0
    \ .
    \label{eq:EOM_psi_order_2}
\end{align}
The EOMs contain new oscillatory source terms induced by the nonlinear interactions through $\varphi_1$ and $\psi_1$.
Substituting  Eqs.~\eqref{eq:solution_varphi_order_1} and \eqref{eq:solution_psi_order_1} into  Eqs.~\eqref{eq:EOM_varphi_order_2} and \eqref{eq:EOM_psi_order_2}, we can rewrite them as
\begin{align}
    \partial_t^2{\varphi}_2 + 
    \varphi_2
    &=
    -\frac{3}{8} \left[
        A^2 e^{-2i t} + 2|A|^2 
        + A^{\ast 2}e^{2i t}
    \right]
    -\frac{1}{4} \mu^2 \left[
        B^2 e^{-2i \mu t} + 2|B|^2 + B^{\ast 2}e^{2i \mu t}
    \right]
    , 
    \\
    \partial_t^2{\psi}_2 + \mu^2 \psi_2
    &=
    -\frac{1}{2} \mu^2 \left[
        A B e^{-i(1+\mu)t}
        + A B^{\ast} e^{-i(1-\mu)t}
        + A^{\ast} B e^{i(1-\mu)t}
        + A^{\ast} B^{\ast} e^{i(1+\mu)t}
    \right]
    \ .
\end{align}
Solving these equations, we obtain
\begin{align}
    \varphi_2
    &=
    \frac{1}{8} \biggl[
        A^2 e^{-2i t} - 6|A|^2 + A^{\ast 2}e^{2i t}
        - \frac{ 2B^2\mu^2e^{-2i\mu t} }{ 1 - 4\mu^2 }
        - 4|B|^2 \mu^2
        - \frac{ 2B^{\ast 2}\mu^2 e^{2i\mu t} }{ 1 - 4\mu^2 }
    \biggr]
        \notag \\
        &~~~~~~~~
        + C_1 \cos t + C_2 \sin t
    \ ,
    \label{eq: phi2 solution}
    \\
    \psi_2
    &=
    \frac{1}{ 2(1 - 4\mu^2) }
    \biggl[
        A^{\ast} \mu^2 e^{it} \left\{
            B e^{-i\mu t} (1+2\mu) + B^{\ast}e^{i\mu t}(1-2\mu)
        \right\}
        \notag \\
        &~~~~~~~~~~~~~~~~~~~
        + A \mu^2 e^{-it} \left\{
            B e^{-i\mu t} (1-2\mu) + B^{\ast}e^{i\mu t}(1+2\mu)
        \right\}
    \biggr]
    \notag \\
    & ~~~ + C_3 \cos (\mu t) + C_4 \sin (\mu t)
    \ ,
\end{align}
where $C_1, \ldots, C_4$ are integration constants with respect to the fast time \(t\). Their dependence on the slow variables \(\tau\) and \(\rho\) is determined only at higher order, and is irrelevant for the leading-order oscillon profiles considered below.

When $m_{\phi}=2m_{\psi}$, i.e., $\mu= 1/2$, the denominator vanishes, and $\varphi_2$ and $\psi_2$ diverge, which indicates a resonance and suggests an instability.
For $m_{\phi} \neq 2m_{\psi}$, the solutions remain oscillatory and stable and do not exhibit any resonant amplification.
Therefore, the stability of multi-field oscillons depends sensitively on the mass ratio.

\subsubsection{The third order \texorpdfstring{$\mathcal{O}(\epsilon^3)$}{}}

The EOMs for the third order $\mathcal{O}(\epsilon^3)$ are given by
\begin{align}
    \partial_t^2\varphi_3+\varphi_3
    &=
    \left( \partial_{\rho}^2+\frac{d-1}{\rho}\partial_{\rho} \right)\varphi_1
    - 2 \partial_t \partial_{\tau} \varphi_1
    - \frac{1}{2} \varphi_1(\varphi_1^2+6\varphi_2)
    - \mu^2 (\varphi_1 \psi_1^2+2\psi_1 \psi_2) 
    \ ,
    \\
    \partial_t^2 \psi_3 + \mu^2 \psi_3
    &=
    \left( \partial_{\rho}^2 + \frac{d-1}{\rho}\partial_{\rho} \right) \psi_1
    - 2 \partial_t \partial_{\tau} \psi_1
    - \mu^2 ( \varphi_1^2 \psi_1 + 2\varphi_2 \psi_1 + 2\varphi_1 \psi_2 )
    \ .
\end{align}
Then, we substitute the solutions for $\mathcal{O}(\epsilon)$ and $\mathcal{O}(\epsilon^2)$ and require that secular terms vanish.
Then, the time evolution of the amplitude functions $A(\tau,\rho)$ and $B(\tau,\rho)$ should follow
\begin{align}
    \left( \partial_{\rho}^2 + \frac{d-1}{\rho}\partial_{\rho}\right)A
    + 2i \partial_{\tau}A
    + \frac{1}{2(1 - 4\mu^2)} \left[
        3 A |A|^2 ( 1-4\mu^2 )
        +  2 \mu^2 A |B|^2 ( 1 - 6\mu^2 )
    \right]
    &=
    0
    \ , 
    \label{eq:EOM_A}
    \\
    \left( \partial_{\rho}^2+\frac{d-1}{\rho}\partial_{\rho} \right)B
    + 2i \mu \partial_{\tau}B
    + \frac{\mu^2}{2 ( 1 - 4\mu^2 )}
    \left[
        2 |A|^2 B ( 1 - 6\mu^2 )
        + \mu^2 B |B|^2 ( 3 - 8\mu^2 )
    \right]
    &=
    0
    \ ,
    \label{eq:EOM_B}
\end{align}
where we assumed the non-resonant case $\mu \neq 1/2$.

By adopting the ansatzes of the separable form, 
\begin{align}
    A(\tau,\rho)
    =
    a(\rho)e^{i\omega_{\varphi} \tau}
    \ , \quad 
    B(\tau,\rho)
    =
    b(\rho)e^{i\omega_{\psi} \tau}
    \ ,
\end{align}
where $\omega_{\varphi}$ and $\omega_{\psi}$ are constants, we obtain the differential equations for the radial profiles $a(\rho)$ and $b(\rho)$ as
\begin{align}
    & \frac{\mathrm{d}^2a}{\mathrm{d}\rho^2}
    + \frac{d-1}{\rho} \frac{\mathrm{d}a}{\mathrm{d}\rho}
    - 2 \omega_{\varphi} a
    + \frac{1}{2(1 - 4 \mu^2)} \left[
        3  ( 1 - 4 \mu^2 ) a |a|^2
        + 2 \mu^2 ( 1 - 6 \mu^2 ) a |b|^2
    \right]
    =
    0
    \ ,
    \label{eq:EOM_a}
    \\
    & \frac{\mathrm{d}^2b}{\mathrm{d}\rho^2}
    + \frac{d-1}{\rho} \frac{\mathrm{d}b}{\mathrm{d}\rho}
    - 2 \mu \omega_{\psi} b
    +\frac{\mu^2}{2 (1 - 4 \mu^2)}
    \left[
        \mu^2 ( 3 - 8 \mu^2) b |b|^2
        + 2 (1 - 6 \mu^2) |a|^2 b 
    \right]
    =
    0
    \ .
    \label{eq:EOM_b}
\end{align}
In a similar way to the single-field case, the complex phases of $a$ and $b$ become constant respectively under the boundary condition below, Eq.~\eqref{eq: a b boundary}. 
Thus, in the following, we set $a$ and $b$ to be real without loss of generality.
These equations can be rewritten using an effective potential $U_\mathrm{eff}$ as
\begin{align}
    & \frac{\mathrm{d}^2a}{\mathrm{d}\rho^2}
    + \frac{d-1}{\rho} \frac{\mathrm{d}a}{\mathrm{d}\rho}
    + \frac{\partial U_\mathrm{eff}}{\partial a}
    =
    0
    \ ,
    \\
    & \frac{\mathrm{d}^2b}{\mathrm{d}\rho^2}
    + \frac{d-1}{\rho} \frac{\mathrm{d}b}{\mathrm{d}\rho}
    + \frac{\partial U_\mathrm{eff}}{\partial b}
    =
    0
    \ .
\end{align}
with 
\begin{align}
    U_\mathrm{eff}(a,b)
    \equiv &
    - \omega_\varphi a^2
    - \mu \omega_\psi b^2
    + \frac{1}{2(1 - 4 \mu^2)} \left[ 
        \frac{3}{4} ( 1 - 4 \mu^2 ) a^4
        + \mu^2 ( 1 - 6 \mu^2 ) a^2 b^2
        + \frac{\mu^4}{4} ( 3 - 8 \mu^2 ) b^4
    \right]
    \ .
    \label{eq: U_eff}
\end{align}
To ensure regular solutions with finite energy, we impose the boundary conditions,
\begin{equation}
\begin{gathered}
    \frac{da}{d\rho}(\rho = 0) 
    =
    \frac{db}{d\rho}(\rho = 0)
    =
    0
    \ ,
    \\
	a(\rho \to \infty)
    =
    b(\rho \to \infty)
    =
    0
    \ .
    \label{eq: a b boundary}
\end{gathered}
\end{equation}

By solving Eqs.~\eqref{eq:EOM_a} and \eqref{eq:EOM_b} under the boundary conditions \eqref{eq: a b boundary}, we derive multi-field oscillons in the form of
\begin{align}
    \varphi(t,r)
    &=
    \epsilon a(\epsilon r) \cos \left[ 
        \left( m_{\phi} - \epsilon^2\omega_{\varphi} \right) t
    \right]
    + \mathcal{O}(\epsilon^2)
    \ ,
    \label{eq: phi oscillon}
    \\
    \psi(t,r)
    &=
    \epsilon b(\epsilon r) \cos \left[ 
        \left( m_{\psi} - \epsilon^2\omega_{\psi} \right) t
    \right] 
    + \mathcal{O}(\epsilon^2)
    \ ,
    \label{eq: psi oscillon}
\end{align}
where we write the variables in the original dimensionful parameters, and in particular, $\omega_\varphi$ and $\omega_\psi$ here are obtained by multiplying $m_\phi$ with their original definition.
These solutions represent spatially localized oscillons with corrections of order $\mathcal{O}(\epsilon^2)$, where the fields $\varphi$ and $\psi$ oscillate with frequencies shifted from their bare mass by small corrections.

The functions, $a$ and $b$, are determined by fixing $\mu$, $\omega_\varphi$, and $\omega_\psi$.
When we translate $a$ and $b$ to the physical fields, $\varphi$ and $\psi$, we have another free small parameter $\epsilon$.
As in the single-field case, there is a degeneracy between $\epsilon$ and $\omega$'s.
From the structure of Eqs.~\eqref{eq:EOM_a} and \eqref{eq:EOM_b}, we find 
\begin{equation}
\begin{aligned}
    a(\rho; \mu, \alpha^2 \omega_\varphi, \alpha^2 \omega_\psi)
    &=
    \alpha \, a(\alpha \rho; \mu, \omega_\varphi, \omega_\psi)
    \ ,
    \\
    b(\rho; \mu, \alpha^2 \omega_\varphi, \alpha^2 \omega_\psi)
    &=
    \alpha \, b(\alpha \rho; \mu, \omega_\varphi, \omega_\psi)
    \ ,
\end{aligned}
\end{equation}
where $\alpha$ is again a rescaling parameter, and $a(\rho; \mu, \omega_\varphi, \omega_\psi)$ denotes the solution of $a$ for $(\mu, \omega_\varphi, \omega_\psi)$, and the same applies to $b$.
Thus, a simultaneous rescaling of $\omega_\varphi$ and $\omega_\psi$ can be compensated by a rescaling of $\epsilon$.
By taking advantage of this degeneracy, we can fix $\omega_\varphi = m_\phi$ and characterize the physical oscillon configurations by the remaining two parameters, $\epsilon$ and $\omega_\psi$.
Therefore, compared with the single-field case, we have one additional parameter.

\subsection{Single-field limit}
\label{subsec: single field limit}

Before considering multi-field oscillons, we investigate the existence of one-field oscillons.
As we see below, the results in the single-field system in Sec.~\ref{sec:oscillon} can be reproduced as special setups of the two-field system.

\subsubsection{Case of \texorpdfstring{$b=0$}{}}

First, we set $b = 0$, i.e., $\psi_1 = 0$, by hand.
Note that $b = 0$ satisfies the differential equation~\eqref{eq:EOM_b} and boundary condition~\eqref{eq: a b boundary}.
In this case, Eq.~\eqref{eq:EOM_a} is reduced to
\begin{align}
    \frac{\mathrm{d}^2a}{\mathrm{d}\rho^2}
    + \frac{d-1}{\rho} \frac{\mathrm{d}a}{\mathrm{d}\rho}
    - 2 \omega_{\varphi} a
    + \frac{3}{2} a^3
    =
    0
    \ ,
\end{align}
which allows the existence of oscillon solutions and is coincident with Eq.~\eqref{eq: a EoM integrate psi}.

\subsubsection{Case of \texorpdfstring{$a=0$}{}}

Next, we set $a = 0$, which satisfies Eqs.~\eqref{eq:EOM_a} and \eqref{eq: a b boundary}.
Then, we obtain
\begin{align}
    \frac{\mathrm{d}^2b}{\mathrm{d}\rho^2}
    + \frac{d-1}{\rho} \frac{\mathrm{d}b}{\mathrm{d}\rho}
    - 2 \omega_{\psi} \mu b
    + \frac{\mu^4(3 - 8\mu^2)}{2 (1 - 4\mu^2)}b^3
    =
    0
    \ ,
\end{align}
which coincides with Eq.~\eqref{eq: b EoM integrate phi} in the limit of $\mu \ll 1$.
While $\psi$ itself has no quartic self-coupling in Eq.~\eqref{eq:expanded_potential_real_FLS}, the effective potential for $b$ has a quartic term with a negative coefficient.
This term can be understood to come from the exchange of $\varphi$ via the interaction term $\propto \varphi \psi^2$.
For this differential equation, oscillon solutions exist when the coefficient of the cubic term in the left-hand side is positive, requiring
\begin{align}
    m_{\psi} < \frac{1}{2}m_{\phi}
    \text{~~or~~}
    m_\psi> \frac{\sqrt{6}}{4}m_{\phi}
    \simeq 0.61 m_{\phi} 
    \ .
\end{align}

In this case, we set $a = 0$, and thus $\varphi$ vanishes at the leading order.
However, it is induced at second order as seen in Eq.~\eqref{eq: phi2 solution}.
Here, $\varphi$ is induced through the $\varphi \psi^2$ term and thus oscillates with an angular frequency of $2 m_\psi$.
This solution corresponds to the excited-state solution found in Ref.~\cite{Gleiser:2011xj}.

\subsection{Multi-field oscillons}
\label{subsec: multi field oscillons}

Next, we consider multi-field oscillons in which both $\varphi$ and $\psi$ (or equivalently, $a$ and $b$) have nonzero field values.

First, we investigate the condition for oscillon solutions to exist analogously to the single-field case.
We can understand Eqs.~\eqref{eq:EOM_a} and \eqref{eq:EOM_b} as the EOMs for two homogeneous scalar fields with the effective potential $U_\mathrm{eff}$ in an expanding universe, if we regard $\rho$ as a time coordinate.
Since $U_\mathrm{eff}$ is an even function of $a$ and $b$, we find
\begin{align}
    \frac{\partial U_\mathrm{eff}}{\partial a}(0,b)
    =
    \frac{\partial U_\mathrm{eff}}{\partial b}(a,0)
    =
    0
    \ .
    \label{eq: Ueff is even}
\end{align}
Thus, if $a = 0$ at $\rho = 0$, then $a$ remains zero throughout space, which also holds for $b$.
This corresponds to the single-field oscillons discussed above.

On the other hand, multi-field oscillon solutions correspond to the solutions that start from nonzero $a$ and $b$ with vanishing velocity at $\rho = 0$ and converge to $a = b = 0$ as $\rho \to \infty$.
To estimate the condition for such solutions to exist, we define
\begin{align}
    a \equiv X \cos \theta
    \ , \quad 
    b \equiv X \sin \theta
    \ 
\end{align}
with $X \geq 0$, and treat $U_\mathrm{eff}$ as a function of $X$ and $\theta$ as $U_\mathrm{eff}(X,\theta)$. We denote the values of $X$ and $\theta$ at $\rho = 0$ as $X_i$ and $\theta_i$, respectively. 
Due to the evenness of $U_\mathrm{eff}$, we can take $0 \leq \theta_i  \leq \pi/2$ without loss of generality.
Then, multi-field oscillon solutions have $0 < \theta_i < \pi/2$.
We can rewrite Eq.~\eqref{eq: Ueff is even} as
\begin{align}
    \frac{\partial U_\mathrm{eff}}{\partial \theta}(X,0)
    =
    \frac{\partial U_\mathrm{eff}}{\partial \theta}\left(X,\frac{\pi}{2}\right)
    =
    0
    \ .
\end{align}
If $\partial U_\mathrm{eff}/\partial \theta (X,\theta) \neq 0$ for any $X < X_i $ and $0 < \theta < \pi/2$, then $\theta$ rolls down in one direction as $\rho$ increases and does not converge to a certain value for $\rho \to \infty$.%
\footnote{Note that the coefficients of the quadratic terms are negative.} 
This is because $\partial_\theta U_\mathrm{eff}\neq 0$ provides a nonzero torque, so the angular momentum in the $(a,b)$ plane does not vanish as $\rho \to \infty$.
In such a case, we cannot obtain multi-field oscillon solutions. 
On the other hand, if there were extrema in the $\theta$ direction, it would be possible to balance the net angular momentum so that it vanishes as $\rho \to \infty$.
Thus, we can consider the existence of extrema in the $\theta$ direction for $0 < \theta < \pi/2$ as a necessary condition for the existence of multi-field oscillon solutions. 
In particular, in the current setup, there can be only one extremum along the $\theta$ direction within $0 < \theta < \pi/2$.

\begin{figure}[ht]
\centering
\includegraphics[width=8cm]{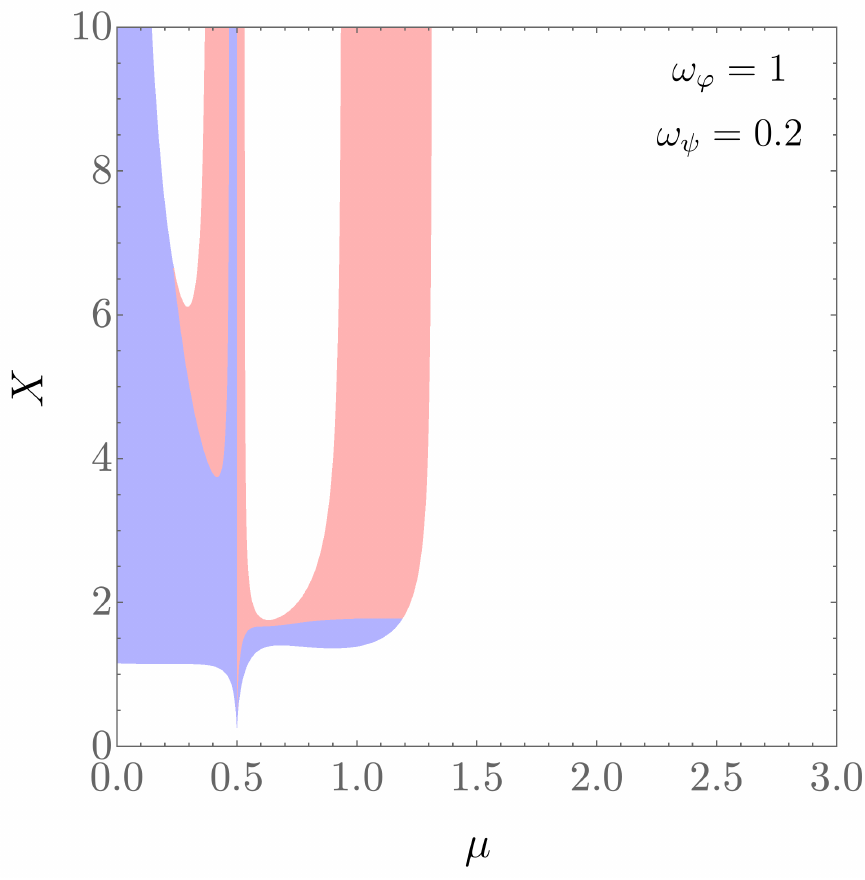}
\includegraphics[width=8cm]{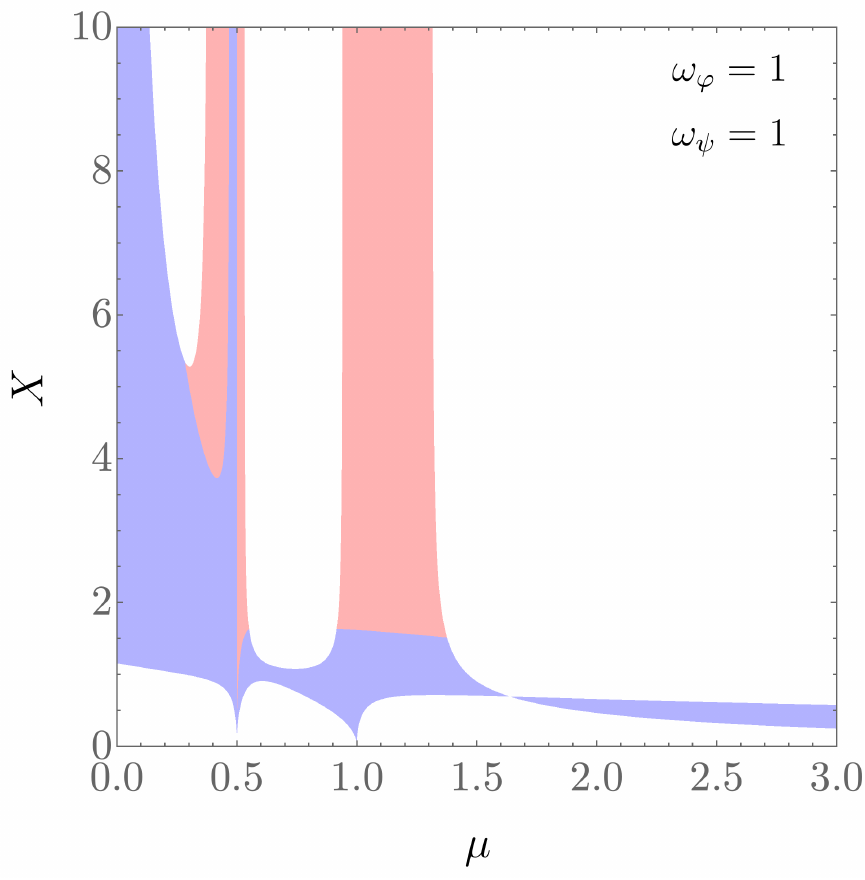}
\\
\includegraphics[width=8cm]{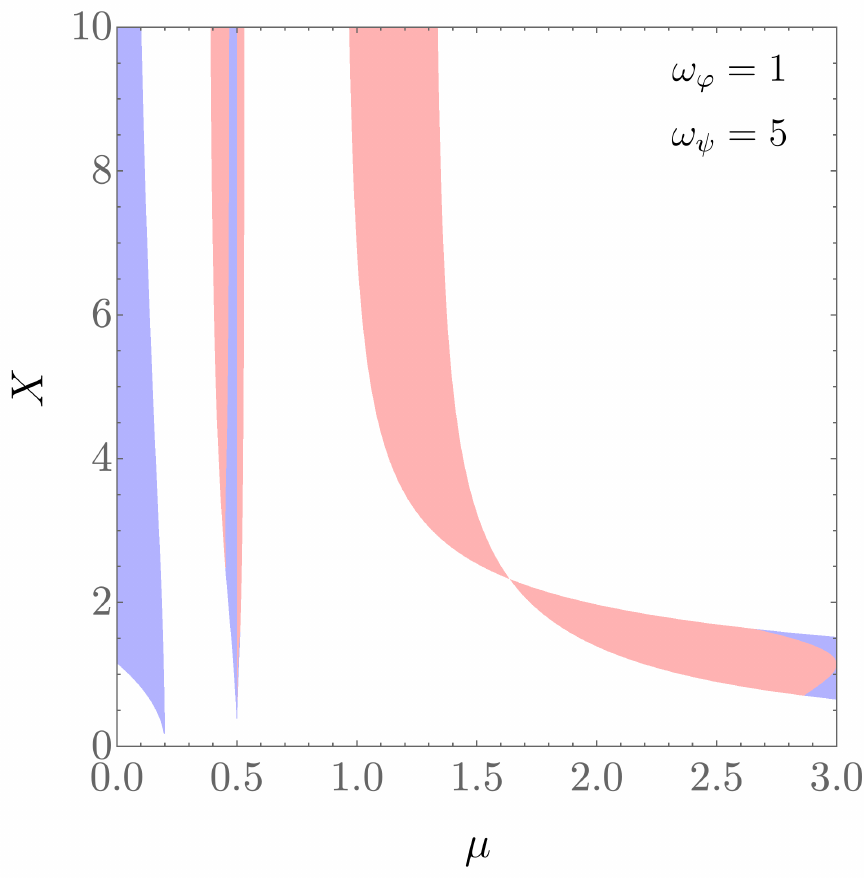}
\caption{
Parameter regions with extrema of $U_\mathrm{eff}$ at $0 < \theta < \pi/2$. 
The red and blue regions denote positive and negative extremum values of $U_\mathrm{eff}$, respectively.
}
\label{fig: eff pot extremum}
\end{figure}

We show in Fig.~\ref{fig: eff pot extremum} the regions of $(\mu,X)$ for which $U_\mathrm{eff}(X,\theta)$ has an extremum along the $\theta$ direction within $0 < \theta < \pi/2$ at fixed $X$.
The red and blue regions denote the positive and negative values of $U_\mathrm{eff}$ at the extrema.
We find that there can be extrema for a wide range of $\mu$ by changing the ratio of $\omega_\varphi$ and $\omega_\psi$. 
Thus, we expect the existence of multi-field oscillon solutions in the parameter region where such extrema exist.
Note that $X_i$ cannot be chosen in the blue region, since $U_\mathrm{eff}(X_i,\theta_i)$ must be positive. We also expect that oscillon solutions are easier to construct when an extremum exists over a broad range of $X$. We also note that, 
for sufficiently small $X$, $U_\mathrm{eff}$ can be approximated by quadratic terms in $a$ and $b$, in which case it varies monotonically along the $\theta$ direction and hence has no extremum.

\begin{figure}[t!]
    \centering
    \includegraphics[width=7.5cm]{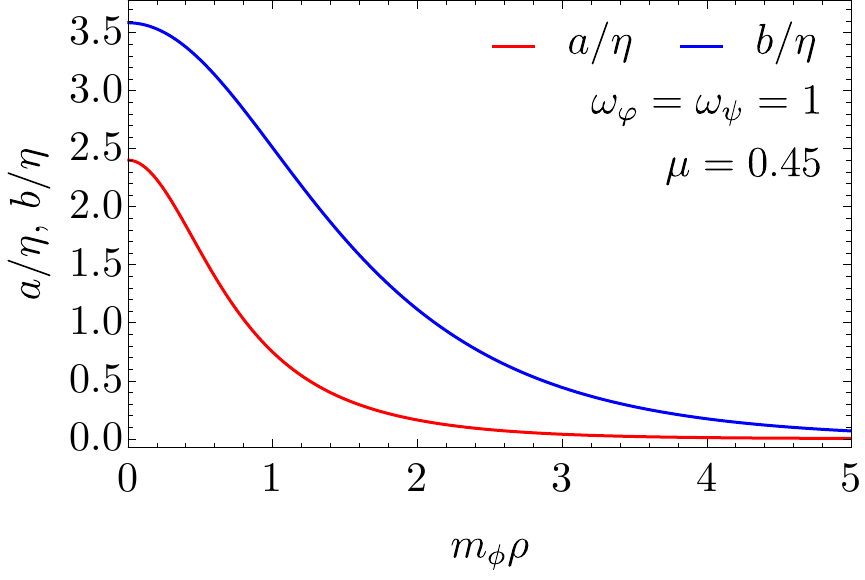}
    \hspace{3mm}
    \includegraphics[width=7.5cm]{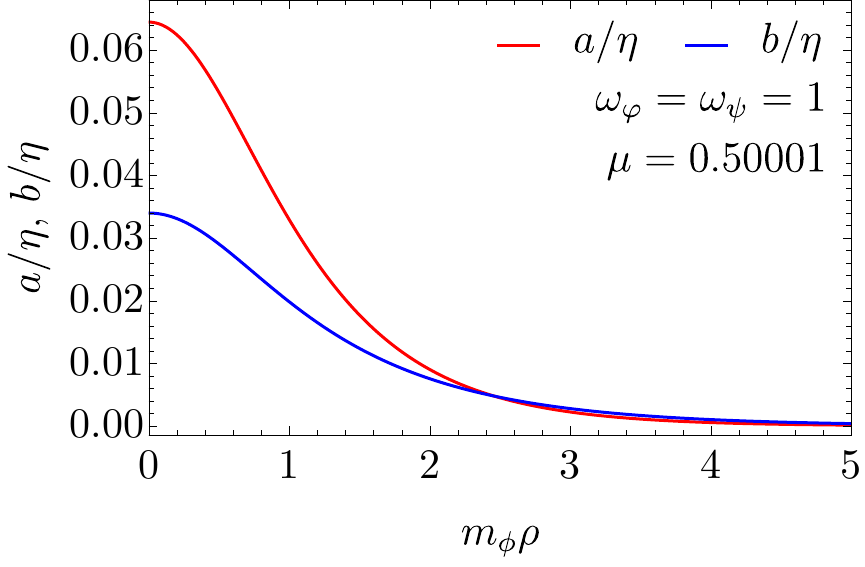}
    \\ \vspace{3mm}
    \includegraphics[width=7.5cm]{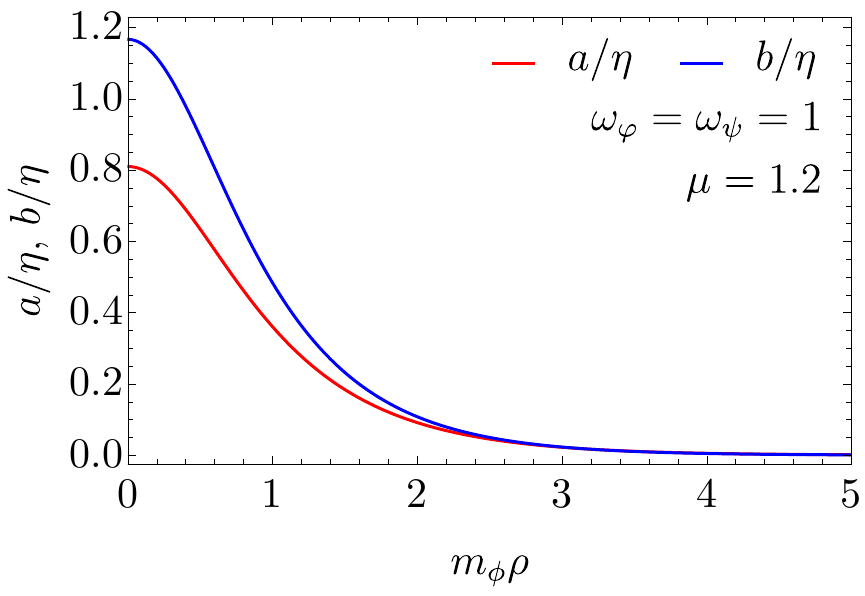}
    \caption{
        Spatial profiles of multi-field oscillons for $\omega_\varphi= \omega_\psi = 1$ and various values of $\mu$.
    }
    \label{fig: multi config}
\end{figure}

The configurations of multi-field oscillons are obtained by solving Eqs.~\eqref{eq:EOM_a} and \eqref{eq:EOM_b} with the boundary conditions~\eqref{eq: a b boundary}.
We show spatial profiles of multi-field oscillons with $d = 1$ in Fig.~\ref{fig: multi config}.
For this analysis, we set $\omega_{\varphi}=\omega_{\psi}=1$ and take three values of $\mu$.
While the multi-field oscillons are found in all the cases, their spatial profiles, absolute amplitudes, and the relative amplitude between $a$ and $b$ depend on the value of $\mu$.
In particular, the amplitude of $a$ can be smaller or larger than $b$ depending on $\mu$.
We also solved Eqs.~\eqref{eq:EOM_a} and \eqref{eq:EOM_b} for $d = 3$ and found qualitatively similar results.

Finally, we discuss whether such configurations are realized in the field dynamics by considering the interaction between $\varphi$ and $\psi$.
Now, we consider small-amplitude oscillons and thus the lowest order term $\propto \varphi \psi^2$ has a dominant effect among the interaction terms (see Eq.~\eqref{eq:expanded_potential_real_FLS}).
When oscillons of $\varphi$ and $\psi$ are colocated, the oscillating $\psi$ induces a linear potential for $\varphi$.
As a result, the oscillation center of $\phi$ is shifted from $\eta$ to a smaller value, leading to a smaller mass of $\varphi$.
This implies that the energy of $\varphi$-oscillons is reduced in the background of oscillating $\psi$.

Regarding the effects on $\psi$-oscillons, there are several possible contributions.
As discussed in Sec.~\ref{subsec: single field limit}, the quartic term in the effective potential for $\psi$, which allows the existence of $\psi$-oscillon solutions, comes from the exchange of $\varphi$.
Thus, the change in the mass of $\varphi$ affects the stability of $\psi$-oscillons.
As noted above, the presence of a $\psi$-oscillon reduces the mass of $\varphi$.
In addition, the effective mass of $\varphi$ decreases in a $\varphi$-oscillon, which appears as a correction to the oscillation frequency of oscillons (see Eq.~\eqref{eq: phi oscillon}).
Conversely, the interaction term $\propto \varphi^2 \psi^2$ increases the $\psi$ mass in the presence of a $\phi$ oscillon, which will induce a repulsive force between $\varphi$ and $\psi$ oscillons.

In this sense, several effects can generate either attractive or repulsive forces between oscillons of $\varphi$ and $\psi$.
If the combined effect yields an attractive force, multi-field oscillons can form dynamically.
While a quantitative analysis of the conditions for the attractive force between $\varphi$ and $\psi$ oscillons is left for future work, in the next section, we will present the realization of multi-field oscillons in certain setups using numerical lattice simulations.

\section{Numerical lattice simulations}
\label{sec:simulation}

In this section, we perform one-dimensional lattice simulations of the real-FLS model.
By comparing the resulting multi-field oscillon configurations with the predictions from the multiple-scale analysis discussed above, we aim to confirm the existence of multi-field oscillons and to assess the validity of the multiple-scale analysis.
In the following, we consider two types of simulation setups.
First, we consider nearly homogeneous initial conditions for $\phi$ and $\psi$ with small fluctuations, and study the formation of multi-field oscillons.
Second, we see the relaxation of oscillon solutions from Gaussian profiles of $\phi$ and $\psi$.
In both cases, we use the dimensionless quantities defined in Eq.~\eqref{eq: dimless quantities} for the lattice simulations and adopt the leapfrog method for the time integration. 

\subsection{Formation of multi-field oscillons}
\label{subsec: lattice random}

In the first setup, we include cosmic expansion, assuming the radiation-dominated universe $a = \sqrt{t/t_\mathrm{in}}$ to suppress the spatial fluctuations at later times.
Then, the EOMs for the dimensionless fields become
\begin{align}
    - \partial_t^2 \phi 
    - \frac{3}{2 t} \partial_t \phi
    + \frac{1}{a^2}\left( \partial_r^2 + \frac{d-1}{r} \partial_r \right) \phi
    - \frac{1}{2} \phi ( \phi^2 - 1)
    - \mu^2 \phi \psi^2
    &=
    0
    \ ,
    \\
    - \partial_t^2 \psi
    - \frac{3}{2 t} \partial_t \psi
    + \frac{1}{a^2}\left( \partial_r^2 + \frac{d-1}{r} \partial_r \right) \psi
    - \mu^2 \phi^2 \psi
    &=
    0
    \ .
\end{align}
We set the initial time to $t_\mathrm{in} = 200$ and evolve the fields until $t_\mathrm{fin} \simeq 10^5$.
At the box boundaries, we impose the periodic boundary condition.
As the initial conditions, we adopt
\begin{align}
    \phi(t_\mathrm{in},x)
    =
    0.30  + \delta \phi(x)
    \ ,
    \quad 
    \psi(t_\mathrm{in},x)
    =
    1.1  + \delta \psi(x)
\end{align}
where $\delta \phi$ and $\delta \psi$ are random noises at each spatial grid point following a uniform distribution over $[-0.001,0.001]$.
For a relatively large amplitude of $\psi$ at the initial time, $\phi$ is temporarily stabilized at the origin in the early stage of the simulation, and then develops domain walls interpolating $\phi = \pm 1$.
When two domain walls collide and annihilate, a large-amplitude excitation of $\phi$ is induced, leading to the formation of an oscillon.
In this sense, we investigate a specific type of oscillon formation in this setup.

\begin{figure}[t!]
\centering
\includegraphics[width=17cm]{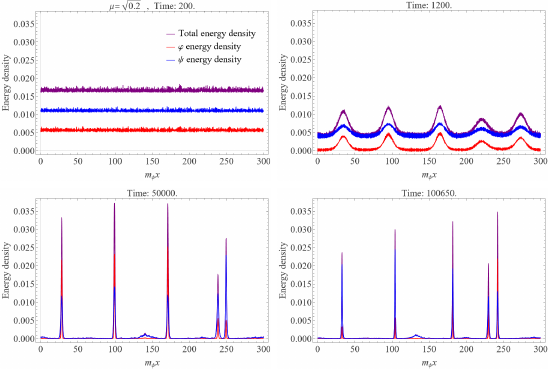}
\caption{
  Snapshots of the energy density from the lattice simulation with random initial conditions for $\mu = \sqrt{0.2}$.
}
\label{fig:1Dsimulation_mu_sqrt0p2_energydensity}
\end{figure}

We show snapshots of energy densities for $\mu=\sqrt{0.2}$ in Fig.~\ref{fig:1Dsimulation_mu_sqrt0p2_energydensity}.
Here, the potential term $\propto \phi^2 \psi^2$ is included in the energy density for $\psi$.
At the initial time (top-left panel), both $\phi$ and $\psi$ have nearly homogeneous energy densities with small fluctuations.
Then, several localized configurations are formed.
In later times (bottom panels), we find five sharp peaks of both $\varphi$ and $\psi$ fields and a few broad peaks of $\psi$.
The former are considered to be multi-field oscillons, while the latter is an oscillon of the $\psi$ field.
We find that the oscillons found in the simulation remain even after $\mathcal{O}(10^4)$ oscillations of the fields.
In multi-field oscillons, the spatial profile of the $\psi$ field is broader than that of the $\varphi$ field, which reflects $\mu < 1$.
In this case, multi-field oscillons are initially sourced by large-amplitude oscillations of $\varphi$, which then induce oscillations of $\psi$ through the interaction terms $\propto \varphi \psi^2$ and $\varphi^2 \psi^2$.
On the other hand, $\psi$-oscillons do not lead to multi-field oscillons because $\psi$ cannot excite $\varphi$ through the interaction due to the small value of $\mu$.

\begin{figure}[t!]
\centering
\includegraphics[width=14cm]{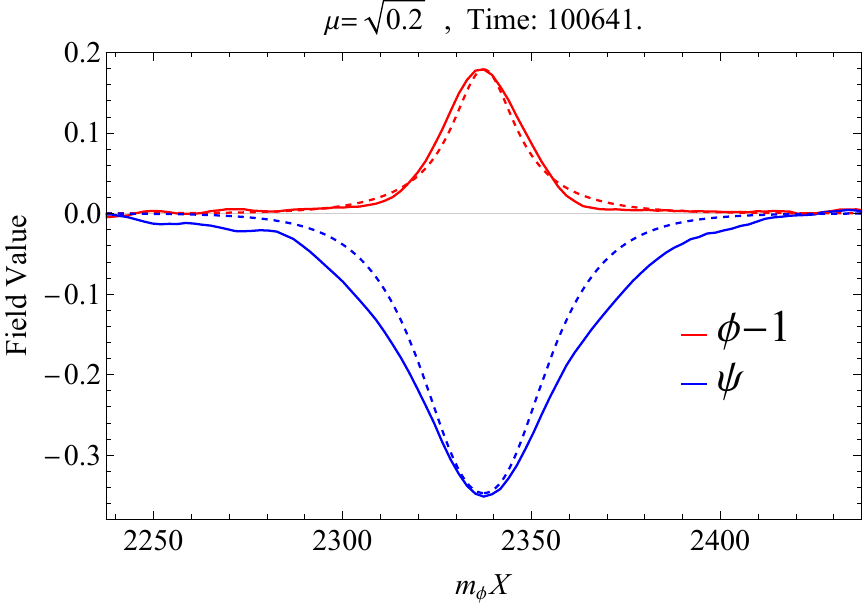}
\caption{
    Snapshot of field configurations from a lattice simulation with random initial conditions.
    The solid lines denote the results of the lattice simulation, while the dashed lines denote the solution obtained from the multiple-scale analysis, where $X \equiv a(t)x$ is the physical length.
}
\label{fig: configuration compare}
\end{figure}

In Fig.~\ref{fig: configuration compare}, we show a snapshot of the field configuration for $\mu=\sqrt{0.2}$ by solid lines.
The dashed lines represent the solution obtained from the multiple-scale analysis, \eqref{eq: phi oscillon} and \eqref{eq: psi oscillon}.
Here, we choose the time such that 
the oscillations of both fields are simultaneously close to their extrema.
By comparing these two results, we determine the parameters to be $\epsilon \simeq 0.0875$ and $\omega_\varphi \simeq 0.385$, with $\omega_\psi$ fixed to $1$.
We find that the analytical solution reproduces the spatial profiles of both fields near the core of the localized configuration with good accuracy, indicating that the multiple-scale analysis captures the essential structure of the configuration.
The slight difference in the spatial radius of the $\psi$ field configuration can be attributed to the deviation from the small-amplitude limit, which is assumed in the multiple-scale analysis.
We also performed this type of simulation and found the formation of multi-field oscillons for different values of the mass ratio, e.g., $\mu = \sqrt{0.3}.$
Thus, multi-field oscillons are found to form in both parameter regimes, $\mu>1/2$ and $\mu<1/2$, while $\mu=1/2$ corresponds to the critical value in the multiple-scale analysis.

\subsection{Relaxation to multi-field oscillons}
\label{subsec: lattice gaussian}

Second, we use Gaussian ansatzes for the initial condition of $\phi$ and $\psi$:
\begin{align}
    \phi(t_\mathrm{in},x)
    =
    1 + \Phi_c \, e^{-\frac{(x-x_c)^2}{R^2}}
    \ , \quad 
    \psi(t_\mathrm{in},x)
    =
    \Psi_c \, e^{-\frac{(x-x_c)^2}{R^2}}
\end{align}
We do not include the cosmic expansion and set the initial time by $t_\mathrm{in} = 0$. 
We set $x_c$ to be the center of the lattice box and impose the absorbing boundary condition at the box boundaries:
\begin{equation}
\begin{aligned}
    &\partial_t^2 \chi
    - \partial_t \partial_x \chi
    + \frac{1}{2} \frac{\partial V}{\partial \chi}
    =
    0
    \quad 
    (\text{Left boundary})
    \ ,
    \\
    &\partial_t^2 \chi
    + \partial_t \partial_x \chi
    + \frac{1}{2} \frac{\partial V}{\partial \chi}
    =
    0
    \quad 
    (\text{Right boundary})
    \ ,
\end{aligned}
\end{equation}
where $\chi = \phi, \psi$.

\begin{figure}[t!]
\centering
\includegraphics[width=14cm]{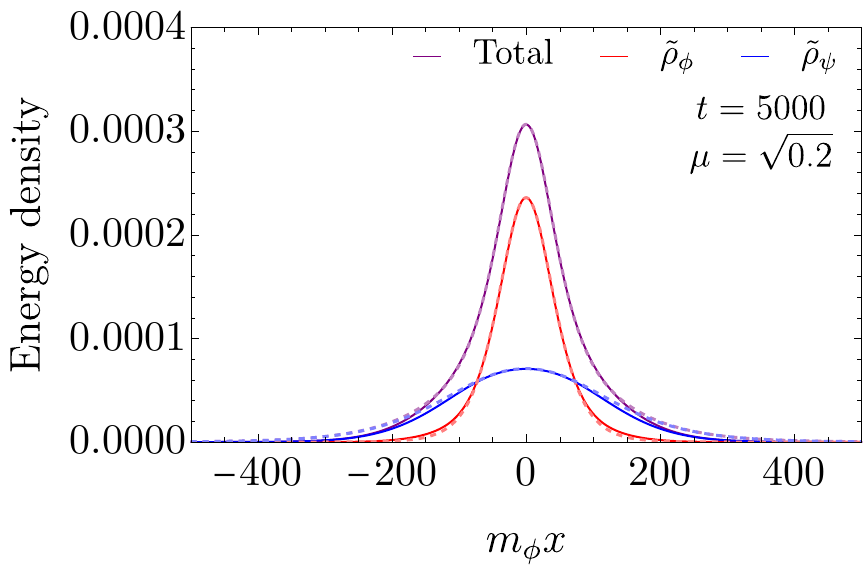}
\caption{
    Snapshots of energy densities in the lattice simulation with the Gaussian initial condition with $\mu = \sqrt{0.2}$.
    The solid lines denote the result of the lattice simulation, and the dashed lines denote the solution from the multiple-scale analysis.
    The energy densities are normalized by $\tilde{\rho}_{\phi/\psi} \equiv \rho_{\phi/\psi}/(m_\phi^2 \eta^2)$.
}
\label{fig: energy Gauss}
\end{figure}

We show a snapshot of the energy densities at $t = 5000$ in Fig.~\ref{fig: energy Gauss}.
Here, we adopt $\mu = \sqrt{0.2}$ and set the initial conditions with $\Phi_c = 0.02$, $\Psi_c = 0.04$, and $R = 100$.
The spatial coordinate is shifted so that $x_c = 0$.
During the time evolution, we find that the energy density of each field continues to vary, indicating an exchange of energy between the two fields, while the total energy remains almost constant.
One possible interpretation is that the oscillon configuration evolves in the two-parameter space of $(\epsilon, \omega_\psi)$ while approximately conserving the total energy.
Such behavior is absent in the single-field case, in which the oscillon is characterized by a single parameter directly related to its energy.
Another possibility is that the numerically obtained configuration includes excited modes, whose contribution to the energy density of each field changes in time.
We do not distinguish these possibilities in the present work, and the observed time evolution may involve both effects.
For comparison, we also show the solution from the multiple-scale analysis with $\mu = \sqrt{0.2}$, $\omega_\varphi = 1$, $\omega_\psi = 0.46$, and $\epsilon = 0.0107$, which are chosen so that the energy densities at the center of the oscillon coincide with those of the lattice result.
While we start from the initial field values of the Gaussian shapes with the same radius for $\phi$ and $\psi$, they relax to the configurations with different radii for the two fields, which are reproduced well by the multiple-scale analysis.
Note that the field amplitudes are much smaller in this case than in the previous one shown in Figs.~\ref{fig:1Dsimulation_mu_sqrt0p2_energydensity} and \ref{fig: configuration compare}, and thus the assumption of $\epsilon \ll 1$ is better satisfied.

\begin{figure}[t!]
\centering
\includegraphics[width=14cm]{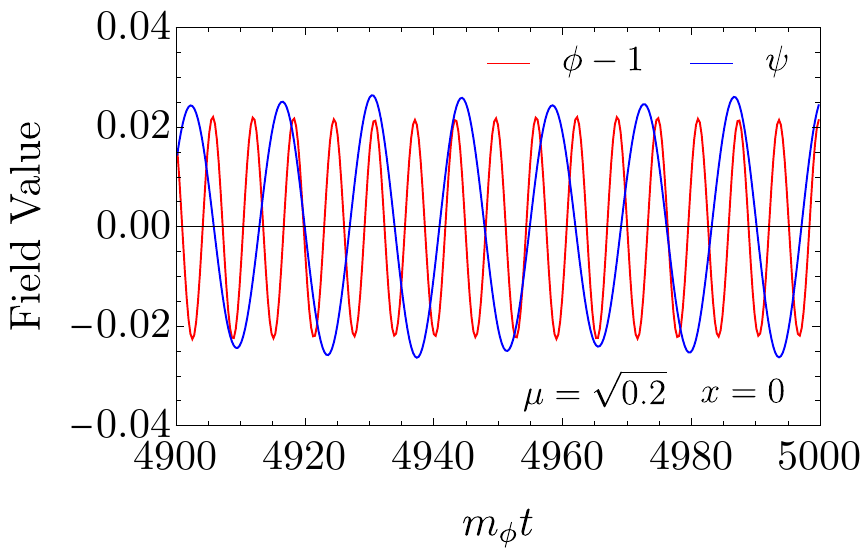}
\caption{
    Time evolution of the fields at the center of the oscillon in the lattice simulation with the Gaussian initial condition with $\mu = \sqrt{0.2}$.
}
\label{fig: center field Gauss}
\end{figure}

We show the time evolution of the field values at $x = 0$ in Fig.~\ref{fig: center field Gauss}.
The two fields exhibit oscillations with distinct frequencies, in agreement with the multiple-scale analysis.
The small corrections to the frequencies are of order $\mathcal{O}(\epsilon^2)$ and are difficult to resolve in the numerical simulation.

While the spatial profiles of the energy densities are well reproduced by the multiple-scale analysis, we find in the lattice simulations that the field oscillations become partially incoherent.
In particular, the oscillation of $\psi$ exhibits different phases in the central and outer regions.
This is because the oscillation amplitude of $\varphi$ varies significantly with spatial position within the oscillon profile.
On the other hand, $\varphi$ exhibits almost coherent oscillations.
The current formalism can describe only coherent oscillations, and to account for such incoherent oscillations, we need to extend the formalism.

We also performed the lattice simulations with the Gaussian initial condition for the three-dimensional space, assuming the spherical symmetry of the system.
In this case, we find that the relaxation of the system takes a longer time and that it is difficult to clearly separate the relaxation from the decay of multi-field oscillons.

\section{Summary and discussion}
\label{sec: summary}

In this study, we investigated the properties of multi-field oscillons in the real-field version of the FLS model.
First, we applied the multiple-scale analysis to the two-field system and showed that multi-field oscillon solutions exist in the real FLS model.
As a result, we find that multi-field oscillon solutions are characterized by two parameters: the ratio of the correction to the oscillation frequencies of the two fields, $\omega_\psi/\omega_\varphi$, which determines the relative amplitude of the two fields at the center of the oscillon, and a small coefficient $\epsilon$, which quantifies the overall amplitude and spatial scale of the oscillon configuration.

Then, we performed lattice simulations and confirmed that multi-field oscillons can be dynamically realized in the real FLS model.
In particular, we examined the formation of multi-field oscillons from random initial conditions and the relaxation of localized initial configurations toward oscillon solutions.
We then validated the multiple-scale analysis by comparing the resulting oscillon solutions with those in the lattice simulations.

While the multiple-scale analysis can be applied to different spatial dimensions, such as $d = 1$ and $d = 3$, we did not perform fully three-dimensional lattice simulations.
In particular, we studied the formation of multi-field oscillons in $d = 1$, and the relaxation process in $d = 1$ and in $d = 3$ under the assumption of spherical symmetry.
One possible extension is to simulate three-dimensional dynamics without assuming any spatial symmetry.
As mentioned above, we found that the relaxation time is longer in three dimensions with spherical symmetry than in one dimension.
Without the spherical symmetry, a more complicated evolution could be observed.

Another is to investigate formation and decay in more realistic settings.
We studied the formation of multi-field oscillons in a setup where domain walls are formed in the early stage, so that we obtain oscillon configurations with suppressed noise.
To assess the cosmological relevance of multi-field oscillons, it is crucial to determine the conditions under which such configurations are dynamically realized in the early universe.
We leave these extensions for future work.

\begin{acknowledgments}
This work is supported by JSPS Core-to-Core Program (grant number: JPJSCCA20200002) (F.T.), JSPS KAKENHI Grant Numbers 20H01894 (F.T.), 20H05851 (F.T.), 23KJ0088(K.M.), 24K17039(K.M.), and 25H02165 (F.T.).
This work is also supported by the World Premier International Research Center Initiative (WPI), MEXT, Japan, and is based upon work from COST Action COSMIC WISPers CA21106, supported by COST (European Cooperation in Science and Technology).
Additional support was provided by the MEXT Promotion of Distinctive Joint Research Center Program (JPMXP0723833165) and the Osaka Metropolitan University Strategic Research Promotion Project (Development of International Research Hubs).
\end{acknowledgments}

\bibliographystyle{JHEP}
\bibliography{Ref}

\end{document}